\documentclass[article]{spie}  

\usepackage{amsmath,amsfonts,amssymb}
\usepackage{graphicx}
\usepackage[colorlinks=true, allcolors=blue]{hyperref}
\usepackage{caption}
\usepackage{subcaption}
\usepackage{array, caption, floatrow, tabularx, makecell, booktabs}%
\usepackage{multicol, multirow}
\usepackage[style=ieee]{biblatex}
\bibliography{references} 
\newfloatcommand{capbtabbox}{table}[][\FBwidth]
\title{Pre-shipment optical characterization of the SCALES instrument}

\author[a]{Isabel J. Kain}
\author[c]{Reni Kupke}
\author[c]{Daren Dillon}
\author[a,c]{R. Deno Stelter}
\author[b]{Rosalie McGurk}
\author[a]{Gwendolin Weber-Stover}
\author[a]{Athira Unni}
\author[b]{Marc Kassis}
\author[b]{Scott Lilley}
\author[d]{Peyton Benac}
\author[a]{Mackenzie Lach}
\author[e]{Arun Surya}
\author[e]{Amirul Hasan}
\author[e]{Ravinder Banyal}
\author[f]{Raquel Martinez}
\author[c]{Nicholas MacDonald}
\author[c]{Will Deich}
\author[c]{Aaron Hunter}
\author[c]{Emily Plume}
\author[c]{Michael Gonzales}
\author[c]{Cristian Rodriguez}
\author[a]{Arjun Kumar}

\author[g]{Cyril Bourgenot}
\author[g]{Paul White}
\author[g]{Juergen Schmoll}
\author[g]{James Wells}
\author[g]{Spencer Davies}
\author[g]{David Bramall}
\author[h]{Dimitri Mawet}
\author[i]{Olivier Absil}
\author[d]{Michael Fitzgerald}
\author[d]{Eric Wang}
\author[a]{Steph Sallum}
\author[a]{Andrew Skemer}
\affil[a]{UC Santa Cruz}
\affil[b]{W. M. Keck Observatory}
\affil[c]{University of California Observatories}
\affil[d]{UCLA}
\affil[e]{Indian Institute for Astrophysics}
\affil[f]{UC Irvine}
\affil[g]{Durham University}
\affil[h]{Caltech}
\affil[i]{University of Liège}

\authorinfo{Corresponding author: Isabel J. Kain (ijkain@ucsc.edu)}

\begin{document} 
\maketitle

\begin{abstract}

The Slicer Combined with an Array of Lenslets for Exoplanet Spectroscopy (SCALES) instrument is a 1-5$\mu$m imager and 2-5$\mu$m integral field spectrograph, currently being commissioned on the Keck II Telescope. SCALES is optimized for exoplanet high-contrast imaging and spectroscopic characterization, and will be sensitive to older, colder exoplanets than existing instrumentation. The 12.3" x 12.3" imaging channel is designed to replicate the capabilities of NIRC2, and the low (R$\sim$35-200, 2.2" x 2.2" FOV) and medium (R$\sim$2500-5000, 0.36" x 0.34" FOV) spectral resolution modes offer new capabilities compared to existing Keck instrumentation. We present preliminary optical performance results from laboratory testing and commissioning of SCALES. 

\end{abstract}

\keywords{Instrumentation, infrared, exoplanets, high-contrast imaging, integral field spectroscopy}

\section{Instrument and testing overview}
\label{sec:instr}

\subsection{Instrument overview}

The Slicer Combined with an Array of Lenslets for Exoplanet Spectroscopy (\href{https://www2.keck.hawaii.edu/inst/scales/}{SCALES}) instrument is a thermal infrared, AO-fed,  high-contrast integral field spectrograph nearing completion for Keck Observatory. Operating at longer wavelengths than existing near-infrared high-contrast IFUs (e.g. GPI, CHARIS, SPHERE), SCALES is sensitive to older and colder exoplanets than have previously been directly imaged from the ground. While SCALES is optimized for exoplanet detection, it will enable a broad range of new science at Keck, including protoplanet characterization, protoplanetary disk mapping, Solar System object monitoring, and characterization of supernova remnants, active galactic nuclei, and more \cite{sallum_slicer_2023}.

SCALES is comprised of a 1-5$\mu$m 12.3\rq\rq x 12.3\rq\rq imaging channel designed to replicate NIRC2’s capabilities, and a low resolution (R$\sim$35-200) and medium resolution (R$\sim$2500-5000) 2-5$\mu$m IFU, offering new capabilities compared to existing Keck instrumentation. The low-resolution IFU uses a lenslet-based design, and the medium-resolution IFU using a novel combined slicer and lenslet array (“slenslit”) design described in \cite{stelter_colors_2021}, \cite{stelter_weighing_2022}, \cite{stelter_scales_2024}. 

There are a large number of individual optics or optical assemblages inside of SCALES between the 4 optical subsystems (see Figure \ref{fig:optics}): coronagraphic masks and non-redundant masks, filters for the imaging channel and IFU, a lenslet array and slicer, dispersive elements in the IFU, and the reflective optics that relay the beam between these subsystems. Each optic impacts instrument performance and, by extension, the science capabilities that SCALES can offer to the Keck user community. 

This paper presents preliminary results from a full optical characterization effort for the SCALES instrument, beginning with pre-ship testing in the UCSC cleanroom and ending once SCALES is fully commissioned at Keck Observatory. Commissioning and optical characterization will be presented more completely in future proceedings and in documentation for observers.


For other pre-ship characterization of the SCALES instrument, see the following SPIE presentations:
\begin{itemize}
\setlength\itemsep{0.05em}
    \item \textbf{SCALES overview:} Paper 14149-19 (Stelter et al. 2026) \cite{richard_d_stelter_update_2026}
    \item \textbf{Slenslet:} Paper 14149-460 (Stelter et al. 2026) \cite{richard_d_stelter_scales_2026}
    \item \textbf{Metasurface scalar vortex coronagraph:} Paper 14154-300 (Palatnick et al. 2026) \cite{skyler_palatnick_metasurface_2026}
    \item \textbf{Non-redundant masks:} Paper 14148-76 (Lach et al. 2026) \cite{mackenzie_r_lach_scales_2026}
    \item \textbf{Spectro-polarimetry:} Paper 14154-392 (Sanchez et al. 2026) \cite{dominic_f_sanchez_infrared_2026}
    \item \textbf{Calibration unit:} Paper 14149-211 (Lach et al. 2026) \cite{mackenzie_r_lach_verifying_2026}
    \item \textbf{Detector software:} Paper 14149-193 (Benac et al. 2026) \cite{peyton_benac_demonstration_2026}
    \item \textbf{Data reduction pipeline:} Paper 14149-181 (Unni et al. 2026) \cite{athira_unni_validating_2026}
    \item \textbf{Spectrograph optomechanics:} Paper 14149-199 (Rodriguez et al. 2026) \cite{cristian_a_rodriguez_scales_2026}
    \item \textbf{Dispersion mechanism optomechanics:} Paper 14154-310 (MacDonald et al. 2026) \cite{nicholas_macdonald_design_2026}
    \item \textbf{Mechanism design \& testing:} Paper 14154-301 (MacDonald et al. 2026) \cite{nicholas_macdonald_scales_2026}
    \item \textbf{Cryostat:} Paper 14149-203 (Gonzalez et al. 2026) \cite{michael_gonzales_design_2026}

\end{itemize}

\subsection{Testing and commissioning overview}

All measurements presented in this paper were taken in a lab setting; most data were taken during the most recent test cooldown of the instrument, CD5 (June 2026). Where more recent data are not available and where no optomechanical or software changes would render data from previous cooldowns inapplicable to the current instrument state, measurements from CD3 (October 2025) and CD4 (January - February 2026) are presented. 

In the absence of a 10-meter telescope on-site at UCSC, light is injected into the instrument by a fiber-fed optical relay called the telescope simulator (TelSim). The TelSim injects light into SCALES from a variety of sources, allowing us to choose flat, point-source, single-mode, or single-wavelength illumination as needed for various optical tests. 

For tests that require precise sampling across wavelength space, we inject light from the McPherson Model 207 scanning monochromator, which is the calibration unit that will be delivered to Keck as part of the instrument package. The SCALES monochromator uses a configuration of internal optics that produce monochromatic light from 1.0-5.2$\mu$m at a minimum theoretical spectral bandwidth of 0.0248 nm or 0.0496 nm, depending on the user's selection of the 300 g/mm or 600 g/mm internal grating.

For testing that requires broadband IR photons, we inject light from the ThorLabs SLS203 globar using either a multi-mode or single-mode IR fiber.

For optical tests that require flood illumination across the entire field of view, we use a coffee warmer set to 60$\deg$C inserted into the collimated region of the TelSim.

\begin{figure}
     \centering
     \begin{subfigure}[b]{0.8\textwidth}
         \centering
         \includegraphics[width=\textwidth]{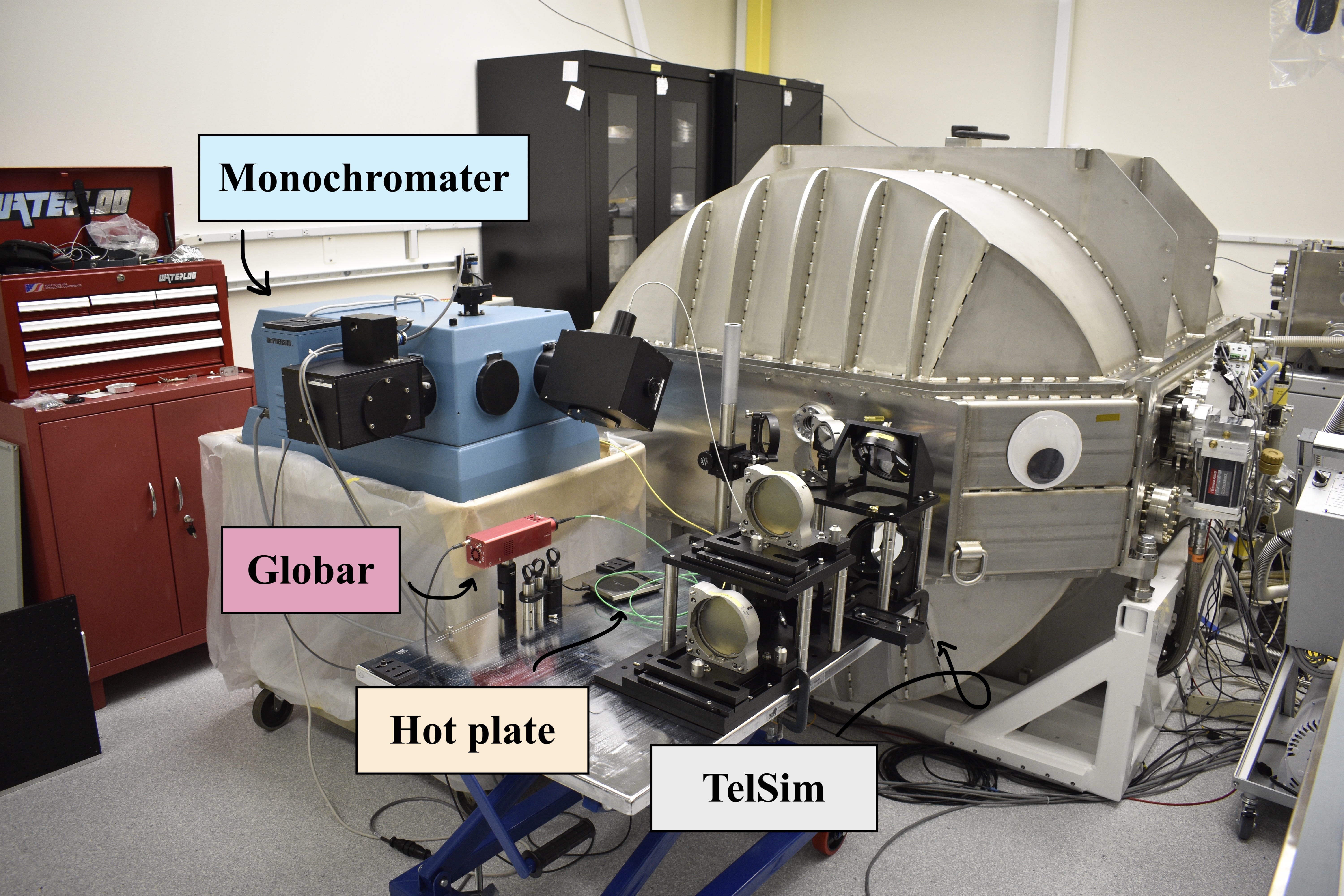}
    \end{subfigure}
        \caption{Instrument and illumination configuration for CD5. The TelSim (bench-mounted optical assembly at center) takes fiber-injected light and relays it into the SCALES entrance window. Light sources used during testing include the monochromater (big blue box), globar (red object mounted to the bench), and hot plate (uninstalled and lying on the TelSim bench behind the globar).}
        \label{fig:inst}
\end{figure}

\section{Results}

\subsection{Throughput}

We present select throughput measurements taken during CD5. Because the IFS filter wheel experienced mechanical problems during this cooldown, throughput testing for IFS of imaging channel filters was not possible. Additionally, scheduling pressures and high-priority mechanism testing took precedence over completeness of optical tests; complete throughput characterization will be presented in future work.

\begin{figure}[ht]
     \centering
     \begin{subfigure}[b]{0.705\textwidth}
         \centering
         \includegraphics[width=\textwidth]{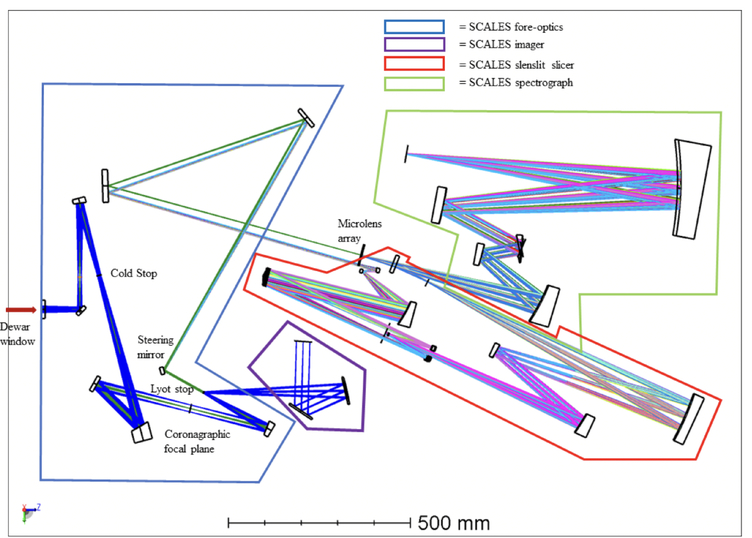}
     \end{subfigure}
     \hfill
     \begin{subfigure}[b]{0.281\textwidth}
         \centering
         \includegraphics[width=\textwidth]{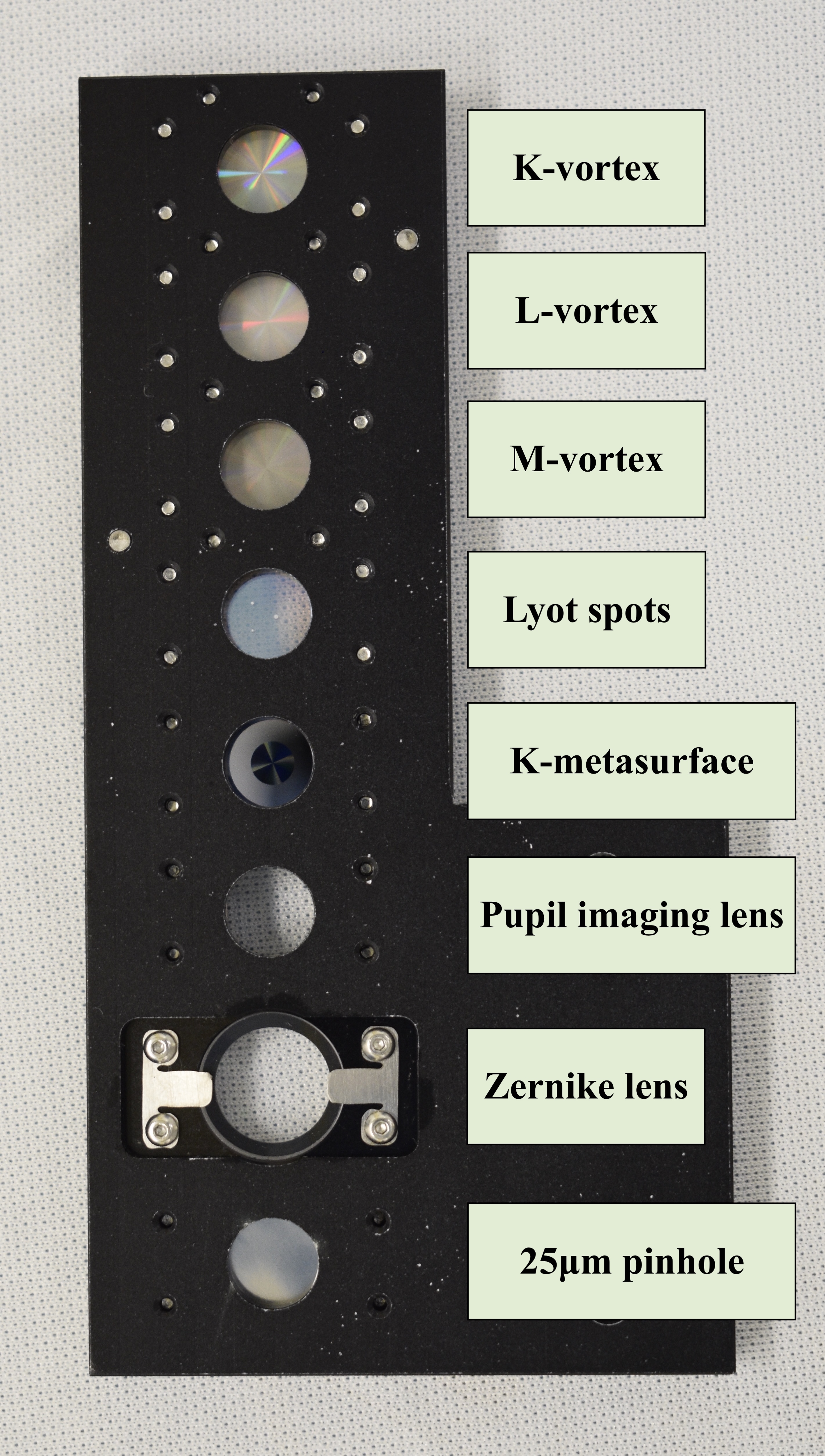}
     \end{subfigure}
        \vspace{0.5mm}
        \caption{Left: SCALES optical diagram. Right: Image of the focal plane coronagraph slide mask holder, with flight optics labeled.}
        \label{fig:optics}
\end{figure}

\subsubsection{Coronagraphs}

SCALES is specialized for exoplanet high-contrast imaging, and is delivered with five coronagraphs. These coronagraphs (and a number of other optics, see Figure \ref{fig:optics}) are mounted onto a linear stage, and are moved vertically into or out of the beam by a mechanism located at a focal plane upstream of both the imager and the IFU.



A vector vortex coronagraph optimized for high-contrast imaging at L' and Ms-band was first installed in NIRC2 in 2015 (\cite{mawet_annular_2005}, \cite{serabyn_w_2017}, \cite{xuan_characterizing_2018}). Vector vortex coronagraphs are still broadly used by current high-contrast instruments, and SCALES will be delivered with three vortex coronagraphs optimized for K, L, and M-band imaging. SCALES' K- and M-vortex coronagraphs were manufactured at Uppsala University, and the L-vortex is inherited from NIRC2. Bowens-Rubin et al. 2025 \cite{bowens-rubin_-sky_2025} found that an absorption feature from the diamond substrate of the NIRC2 L/M vortex coronagraph was imparting a large throughput hit on M-band images, which motivated the manufacturing of a new, much thinner (0.1mm) diamond-substrate M-vortex mask for SCALES.

SCALES will also be shipped with a novel scalar vortex metasurface mask optimized for K-band, which was manufactured at UC Santa Barbara \cite{palatnick_-lab_2026}. This new coronagraph is a first-generation demonstration of scalar vortex coronagraphs, which are designed as high-throughput, polarization-insensitive, achromatic masks for deeper high-contrast imaging sensitivities.

The final coronagraph included in SCALES is a classic Lyot coronagraph with three simple chrome occulting spots arranged in an equilateral triangle on a calcium fluoride substrate. Two of these spots will be aligned with the med-res mode, and one will be aligned with the low-res mode. 

\begin{figure}
     \centering
     \begin{subfigure}[b]{1\textwidth}
         \centering
         \includegraphics[width=\textwidth]{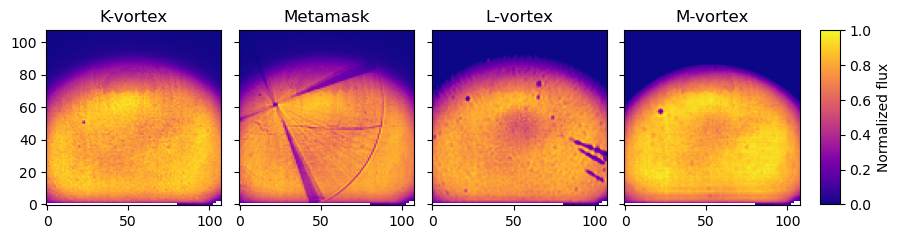}
    \end{subfigure}
        \caption{The four SCALES coronagraphs imaged onto the low-res IFU. The field of view is unevenly illuminated because of misalignment in the TelSim – the circular shape is the footprint of the multimode fiber imaged onto the detector, not the edges of the coronagraph (as can be seen more clearly for the metamask). Only well-illuminated regions are included in analysis. The central occultation spot is visible at the center-left of each image.}
        \label{fig:coron-img}
\end{figure}

Here we present broadband throughput measurements of four of SCALES' coronagraphs: the K- L- and M-band vector vortex coronagraph, and the new K-band metasurface scalar vortex coronagraph. We injected light from the globar and used the low-res IFU (with a mirror in the spectrograph so that the final image is undispersed) to image the coronagraph focal plane, comparing images with the coronagraph slide in the "open" position with images where each coronagraph is inserted into the beam. Figure \ref{fig:coron-img} shows images of each coronagraph, reconstructed by measuring the flux in each undispersed lenslet spot. Because of (now corrected) misalignment in both the coronagraph slide and the TelSim, the masks are unevenly illuminated and are not centered in the FOV. Because of testing of the IFS filter wheels, the imaging channel filter wheels were used as blocking filters and the IFS filter wheels were set to the "open" position, inducing a defocus on the low-res IFU lenslet spots. However, this should not affect the results presented here. The throughput analysis is limited to fully-illuminated regions, and includes the low-throughput center of each mask (such a small region does not significantly impact the average throughput); the central obscuration of each mask is visible in the center-left of each panel of Figure \ref{fig:coron-img}.

\begin{figure}
     \centering
     \begin{subfigure}[b]{1\textwidth}
         \centering
         \includegraphics[width=\textwidth]{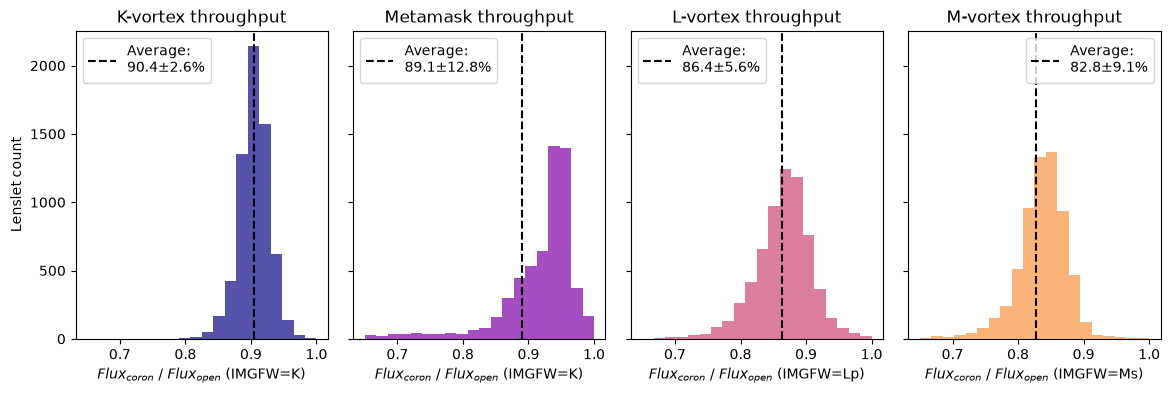}
    \end{subfigure}
        \caption{Throughputs of the four available SCALES coronagraphs as measured during CD5. The throughput of the three vortex coronagraphs are measured across the illuminated portion of the field of view. Each histogram shows the distribution of throughputs measured within an individual lenslet micropupil.} 
        \label{fig:coron-thru}
\end{figure}

Figure \ref{fig:coron-thru} shows the measured throughputs for each mask. We find that all four vortex coronagraphs show excellent throughput of $>$80\% across the board. 

The throughput of the metasurface mask varies spatially due to the charge-4 phase profile, creating a broader throughput distribution across the full face of the mask. We measure a throughput of 89.1\% averaged across the entire clear optic, with low-throughput regions along the charge-4 "spokes" (see Figure \ref{fig:coron-img}) with a typical throughput of 45\%, and a high-throughput peak in the throughput distribution sitting at 95\% throughput. 

Bowens-Rubin et al. 2025 \cite{bowens-rubin_-sky_2025} measured throughputs of the current NIRC2 L/M vortex coronagraph (which was installed in 2015 \cite{serabyn_w_2017}, replacing the $L'$ vortex that SCALES has now inherited) to be 82\% for Lp-band and 57\% for Ms-band. Their Lp-band results are in line with our measured throughput of 86.3\%, and the newly manufactured M-vortex shows enormous improvement, with throughput comparable to the L-vortex. Because our L-vortex coronagraph had a previous life inside of NIRC2, we expect throughput to be minorly degraded by normal wear and tear, as seen in the small specs and scratches along the outer regions in Figure \ref{fig:coron-img}.



\subsubsection{Slicer}

SCALES' IFU channel is comprised of a low-resolution (R$\sim$35-200) and med-resolution (R$\sim$2500-5000) mode. In low-res mode, the SCALES beam is relayed onto the lenslet array, which creates a 103x110 grid of micropupils; these micropupils are then sent through the spectrograph, where the spots are dispersed into "spaxels" that are separated from their neighbors by 108$\mu$m (6 pixels). 

\begin{figure*}
    \centering
    \setkeys{Gin}{width=1.0\textwidth, keepaspectratio}
    \begin{subfigure}{0.99\textwidth}  
        \includegraphics{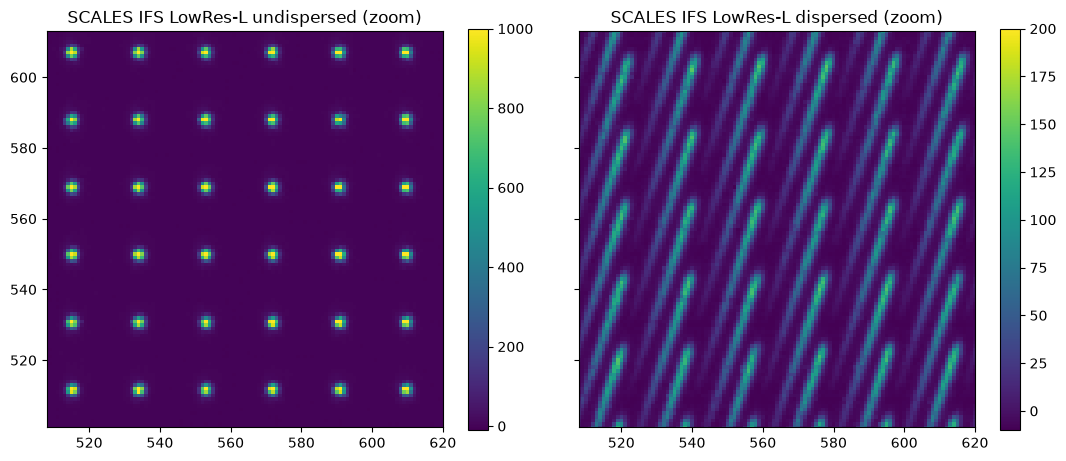}
        \caption{A zoomed-in cutout of lowres L-band data. Left: undispersed lenslet spots. Right: dispersed lenslet spots.}
        \label{fig:example-lowres}
    \end{subfigure}
    
    \begin{subfigure}{1.0\textwidth}  
        \includegraphics{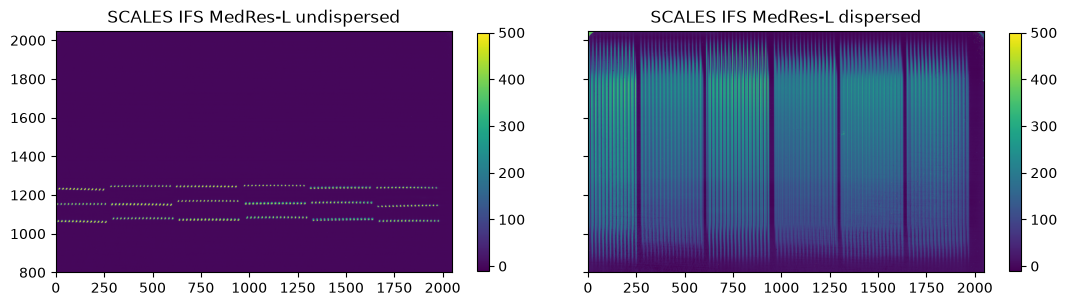}
        \caption{Medres data, with empty regions of the detector cropped for convenience. Left: undispersed lenslet spots. A 17x18 subset of the lenslet array is reimaged by the slicer into a pseudoslit with 88 slightly staggered groups of 17 spots each. Right: dispersed lenslet spots. Note that a blocking filter was used for these measurements, to the wavelength range (and thus the length of the traces) is truncated in this image compared to SCALES' operational configuration. Also note that the leftmost column of spots falls off the detector during CD5, which has now been realigned.}
        \label{fig:example-medres}
    \end{subfigure}
    \caption{Example IFS data, showing the difference between low- and med-res modes.}
    \label{fig:example-data}
\end{figure*}

\begin{figure}[ht]
     \centering
     \begin{subfigure}[b]{0.485\textwidth}
         \centering
         \includegraphics[width=\textwidth]{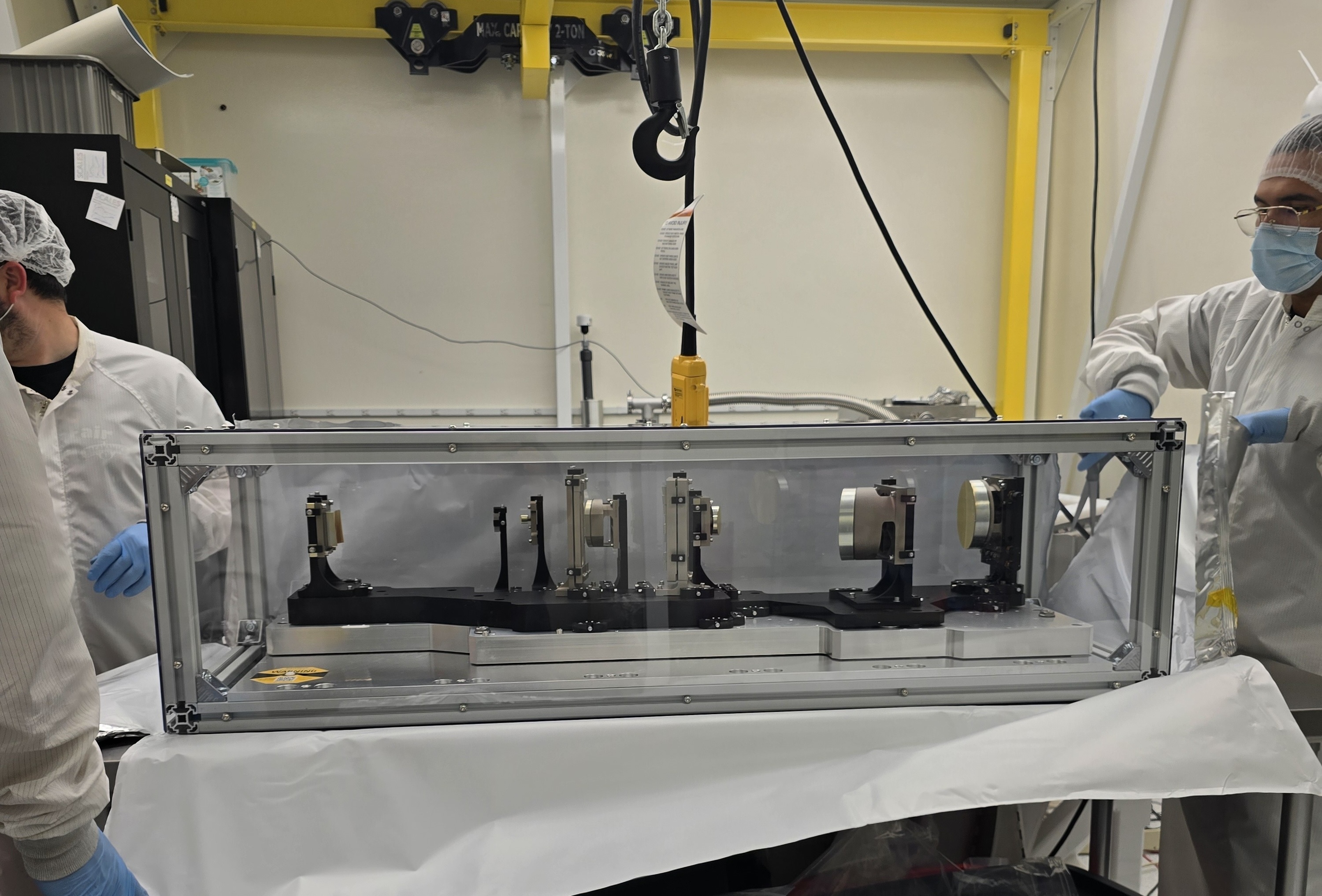}
         \caption{The slicer assembly upon delivery to UCSC.}
     \end{subfigure}
     \hfill
     \begin{subfigure}[b]{0.495\textwidth}
         \centering
         \includegraphics[width=\textwidth]{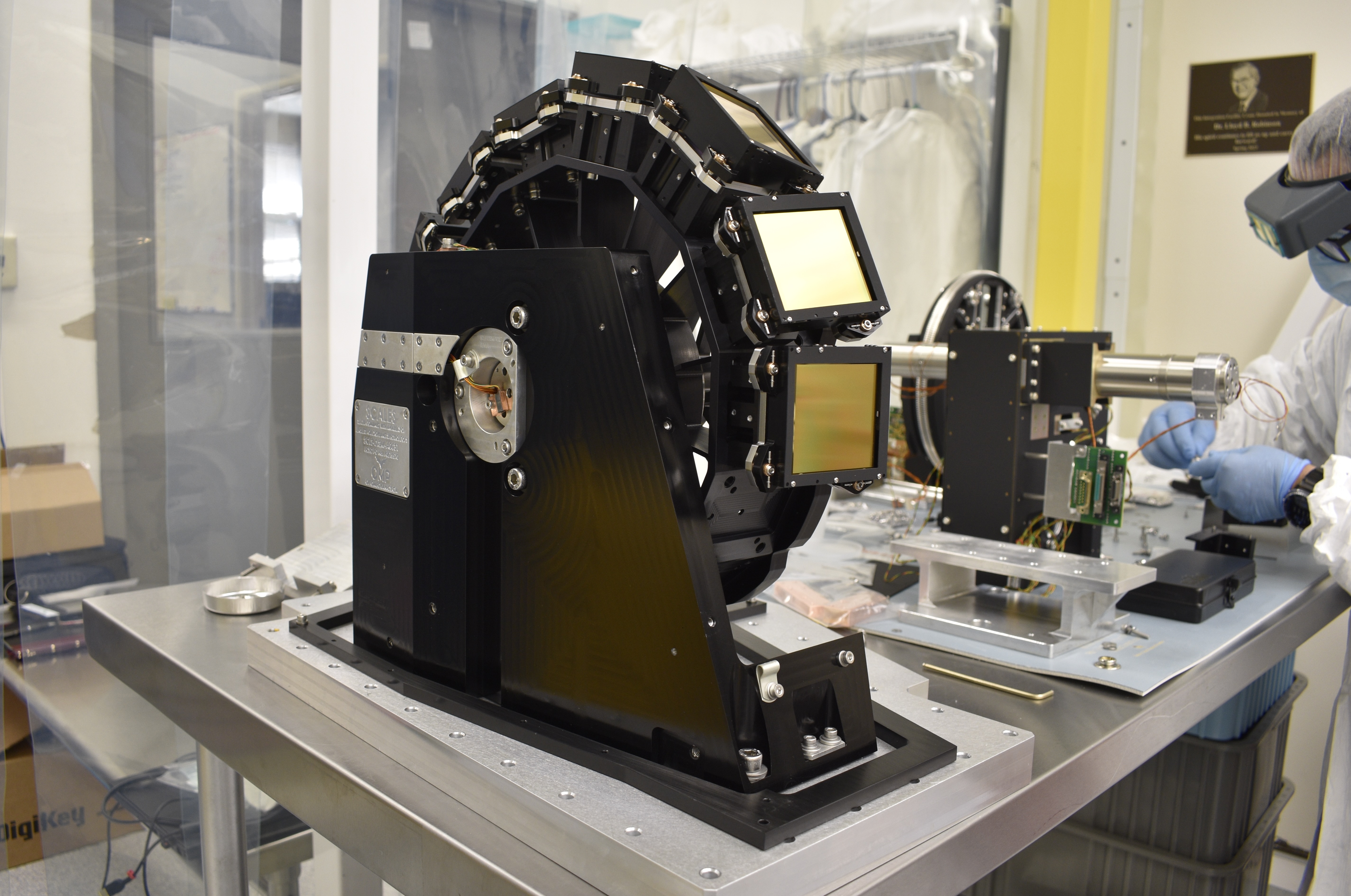}
         \caption{The dispersion carousel.}
     \end{subfigure}
        \vspace{0.5mm}
        \caption{Mid-installation photos from the UCSC cleanroom facility.}
        \label{fig:lab-photos}
\end{figure}

SCALES' med-resolution IFU uses a novel combined lenslet-and-slicer ("slenslet") design. In med-res mode, the lenslet array is masked except for a 17x18 lenslet subregion, which is an on-sky field-of-view of 0.34" x 0.36". This smaller grid of micropupils, instead of being sent straight into the spectrograph, passes through the slicer (pictured in Figure \ref{fig:lab-photos}). The slicer geometrically re-formats the 17x18 grid of micropupils into a staggered 3x102 pseudoslit such that they can be dispersed at much higher resolution without overlap or crosstalk between the resulting spaxels. For two examples of what the low- and med-res data look like, see Figure \ref{fig:example-data}. 

The slicer \cite{richard_d_stelter_scales_2026} adds 10 reflections to the optical surfaces encountered by photons passing through the IFS. The slicer is comprised of 64 individual optics cut into 14 distinct substrates; the optics are single-point diamond-turned RSA 6061 aluminum with a mix of bare and protected gold coatings, which are each expected to impart a $\sim$1\% reduction in throughput.

We measured the broadband throughput of the slicer at L-band (2.8-4.25$\mu$m). We flood-illuminated the entire FOV with the hot plate, positioned a mirror in the dispersion carousel such that the grid of spots was focused undispersed onto the detector, and did aperture photometry on each set of spots. The med-res mode had a moderate defocus error during CD5, which has since been corrected. To compensate for this, we used larger apertures with a radius of 9 pixels for measuring flux across both low- and med-res spots. We measure the throughput of the slicer to be 89.5$\pm$1.6\%. Small adjustments in selected aperture radius vary this result by $\lesssim$0.5\%.

Because of the corrected defocus in the med-res channel (and a number of other more minor changes to SCALES' optical alignment and stray light suppression), this measurement will be repeated during commissioning to verify this result, but this is a useful confirmation that the slicer imparts only a modest throughput hit on the med-res IFU.



\begin{figure}[ht]
     \centering
     \begin{subfigure}[b]{0.8\textwidth}
         \centering
         \includegraphics[width=\textwidth]{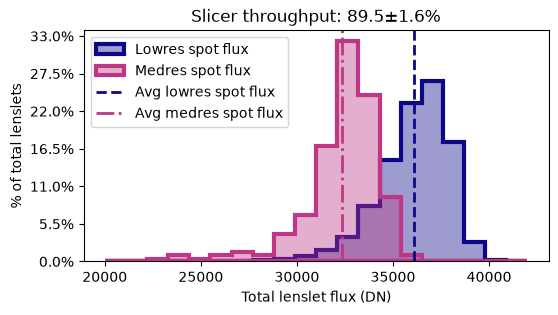}
    \end{subfigure}
        \caption{Broadband throughput of the slicer estimated at L-band (2.8-4.25$\mu$m).}
        \label{fig:slicer-thru}
\end{figure}


\subsubsection{IFS dispersers}

The SCALES IFS has a set of dispersive optics for both low-res and med-res observations: low-resolution (R$\sim$35-200) prisms (K, KL, KLM, Ls, L, M), a KL R$\sim$20 magnesium fluoride Wollaston prism \cite{dominic_f_sanchez_infrared_2026}, and a set of R$\sim$2500-5000 reflective ruled master gratings. The prisms were made by Optimax, the Wollaston prism made by Karl Lambrecht, and the gratings were made by Omega Optical. Here we present throughput measurements of the gratings; a complete analysis of optical throughput on the IFS side will be conducted during commissioning.

\begin{figure}[ht!]
     \centering
     \begin{subfigure}[b]{0.49\textwidth}
         \centering
         \includegraphics[width=1.0\textwidth]{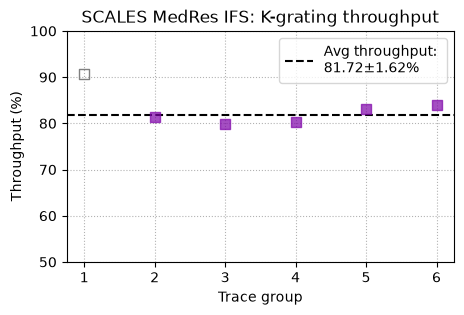}
         \caption{Throughput measured for each 51-trace/spot cluster. The leftmost group of spots partly fall off the detector because of misalignment in the med-res channel during CD5, so this column is thrown out for this analysis.}
     \end{subfigure}
     \hfill
     \begin{subfigure}[b]{0.49\textwidth}
         \centering
         \includegraphics[width=1.0\textwidth]{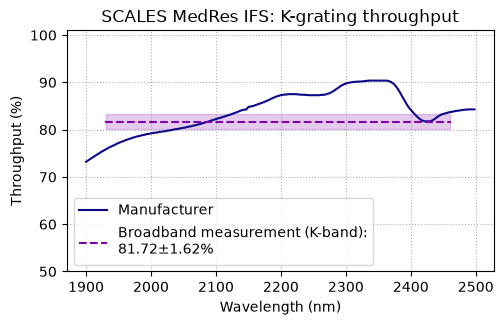}
         \caption{Final broadband throughput measurement of the K-grating (1.93-2.46 $\mu$m) measured inside of SCALES compared to the manufacturer measurement. }
         \vspace{3.8mm}
     \end{subfigure}
        \caption{K-grating throughput as measured inside of SCALES shows good agreement with the manufacturer specifications.}
        \label{fig:k-grating}
\end{figure}

To measure grating throughput, we compared the total flux of the undispersed mirror spots to the dispersed med-res traces when illuminated with the globar. med-res traces are designed to have a center-to-center separation of 6 pixels, with is expected to eliminate crosstalk. However, during CD5 the med-res spots had significant defocus (which has since been aligned out); with this defocus, adjacent traces showed significant overlap, which made extraction of individual traces difficult. To circumvent this, we did aperture photometry on each of the six columns of 51 lenslet traces, and compared these aperture fluxes to the summed fluxes of the corresponding columns of undispersed lenslet spots. Figure \ref{fig:k-grating} shows this throughput measurement for each cluster of spots/traces, and shows that our results are in alignment with the manufacturer specifications.


\subsection{Low-res IFU optical performance}

We present several measurements of optical performance of the SCALES low-resolution IFU. 

The SCALES foreoptics relay the input beam to a focal plane at front surface of the lenslet array (made by Jenoptik), which spatially samples the field. Light passes through each (square) 341x341 $\mu$m silicon lenslet, and are focused into a grid of pupil images; these micropupils are then apertured by a grid of 108$\mu$m diameter circular pinholes, which eliminates crosstalk between adjacent lenslet spots by suppressing the large square diffraction spikes imparted by the lenslet array \cite{peters-limbach_conceptual_2012}. If the IFU is in low-res mode, the lenslet spots are sent directly to the spectrograph, where the beam is collimated, dispersed by a selectable prism in the dispersion carousel, then focused onto the detector. If the IFU is in med-res mode, the lenslet spots take a scenic bypass through the slicer, where the micropupil grid is geometrically reformatted into a pseudoslit, which is then relayed through the spectrograph. For an overview of SCALES' optical layout, see Figure \ref{fig:optics}.

Aberrations imparted on the lenslet spots do not degrade image quality – image quality is finalized at the front surface of the lenslet array, since the this is where the beam is spatially sampled. This has motivated tight surface error requirements for the reflective foreoptics \cite{kain_characterization_2023}. In Section \ref{subsec:image-quality} we present a preliminary assessment of low- and med-res PSFs, though further work will present end-to-end measurements of wavefront error. However, assessment of the lenslet spots themselves is also necessary to characterize instrument performance. While aberrations imparted on the lenslet micropupils do not significantly impact image quality, sufficiently large aberrations downstream of the lenslet array can introduce spectral crosstalk or overlap between adjacent lenslet spots. We characterize lenslet spot quality and geometric distortion in Sections \ref{subsec:optical-quality} and \ref{subsec:distortion}, and measure enslitted energy in Section \ref{subsec:ee}.

\subsubsection{Lenslet spot quality}
\label{subsec:optical-quality}

\begin{figure}[h!]
     \centering
     \begin{subfigure}[b]{0.49\textwidth}
         \centering
         \includegraphics[width=1.0\textwidth]{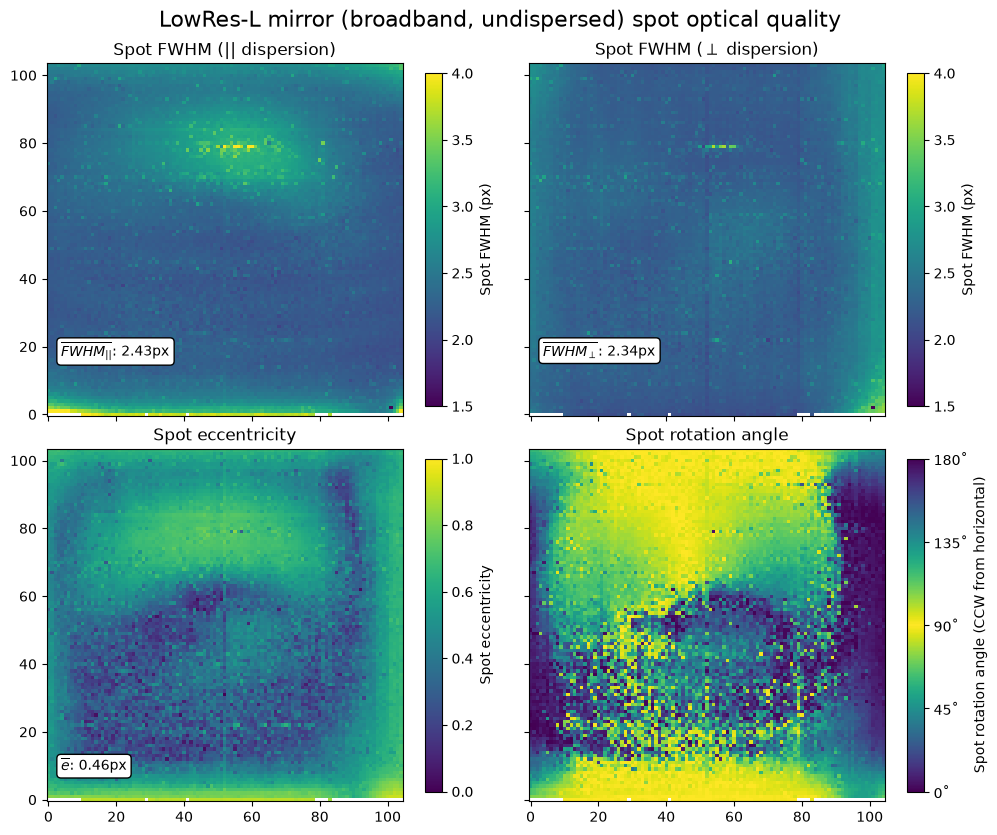}
     \end{subfigure}
     \hfill
     \begin{subfigure}[b]{0.49\textwidth}
         \centering
         \includegraphics[width=1.0\textwidth]{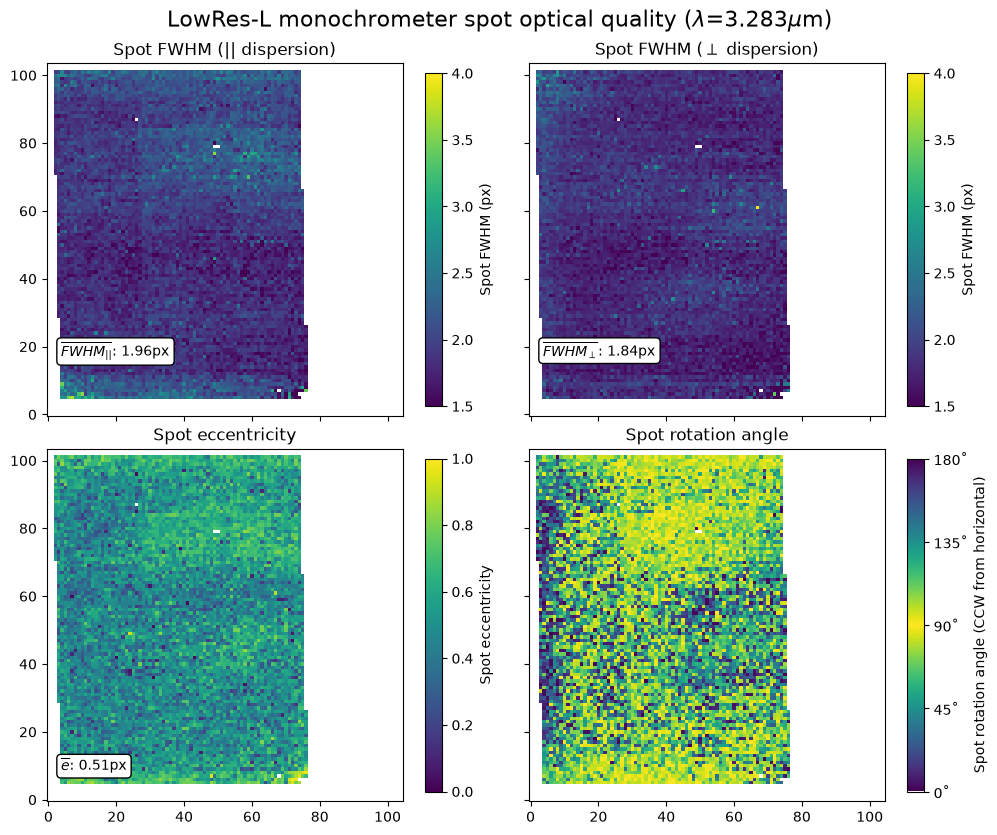}
     \end{subfigure}
        \caption{Measurements of lenslet spot optical quality for the low-res IFU in L-band (2.8-4.25$\mu$m), including spot FWHM parallel and perpendicular to the direction of dispersion, ellipticity, and position angle. Left: optical quality measurements of undispersed mirror spots. Right: optical quality measurements of dispersed single-wavelength ($\lambda$=3.283$\mu$m) spaxels.}
        \label{fig:fwhm}
\end{figure}

In a perfect optical system, the lenslet micropupils will be focused into perfect PSFs. However, the lenslet PSFs are impacted by optical aberrations downstream of the lenslet array. While this has no impact on IFU image quality, and the DRP accounts for spot shape when reconstructing raw IFU images into datacubes, elongation of lenslet spots along the direction of spectral dispersion determines maximum spectral resolution. Elongation of lenslet spots orthogonal to the direction of spectral dispersion may induce crosstalk and impact spatial resolution, which is discussed in Section \ref{subsec:ee}.

To measure the optical quality of the low-res IFU, we fit 2D gaussian profiles to lenslet spots in a low-res L-band dataset taken during CD5. These results, and a comparison between undispersed and dispersed monochromatic spots, are shown in Figure \ref{subsec:optical-quality}.

The FOV of the dispersed monochromater data is truncated because of both uneven illumination from the TelSim and low SNR; the remaining portion of the image is roughly registered with the undispersed lenslet spot data for ease of visual comparison. Certain lenslet spots were poorly handled by the analysis routine (mostly around the edges, with a few in the upper-center of the FOV) because they're significantly dimmer than their neighbors. Edge spots are expected to be lower-signal, but the cluster of dimmer lenslets near the center of the FOV may be better explained otherwise, e.g. manufacturing defects or dust on the surface of the lenslet array. 

Since we approximated the spots with Gaussian instead of Airy profiles, we use the approximation $FWHM_{Gauss} \sim 0.9 \times \lambda \times f/\#$ instead of $FWHM_{Airy} \sim 1.22 \times \lambda \times f/\#$. For a diffraction-limited system, we expected the mirror spots to have a FWHM of 43.22$\mu$m or 2.40 pixels, and the monochromatic spots ($\lambda$=3.283$\mu$m) to have a FWHM of 33.39$\mu$m or 1.85 pixels. Figure \ref{fig:fwhm} shows that spot size orthogonal to the direction of dispersion matches closely with these spot size estimates for a diffraction-limited system. Spots show slight elongation along the direction of dispersion, which will impact spectral resolution; because of the limitations of data taken during CD5, we leave estimation of spectral resolution to future work.

\subsubsection{Lenslet spot distortion}
\label{subsec:distortion}

While lenslet spot distortion is important to characterize for validation of instrument optical performance, it is accounted for by the rectification procedure of the SCALES data reduction pipeline (see Unni et al. 2026 \cite{athira_unni_validating_2026}) – distortion of lenslet spots does not impact overall image quality and should not concern observers. 

\begin{figure*}[h!]
    \centering
    \setkeys{Gin}{width=1.0\textwidth, keepaspectratio}
    \begin{subfigure}{0.99\textwidth}  
        \includegraphics{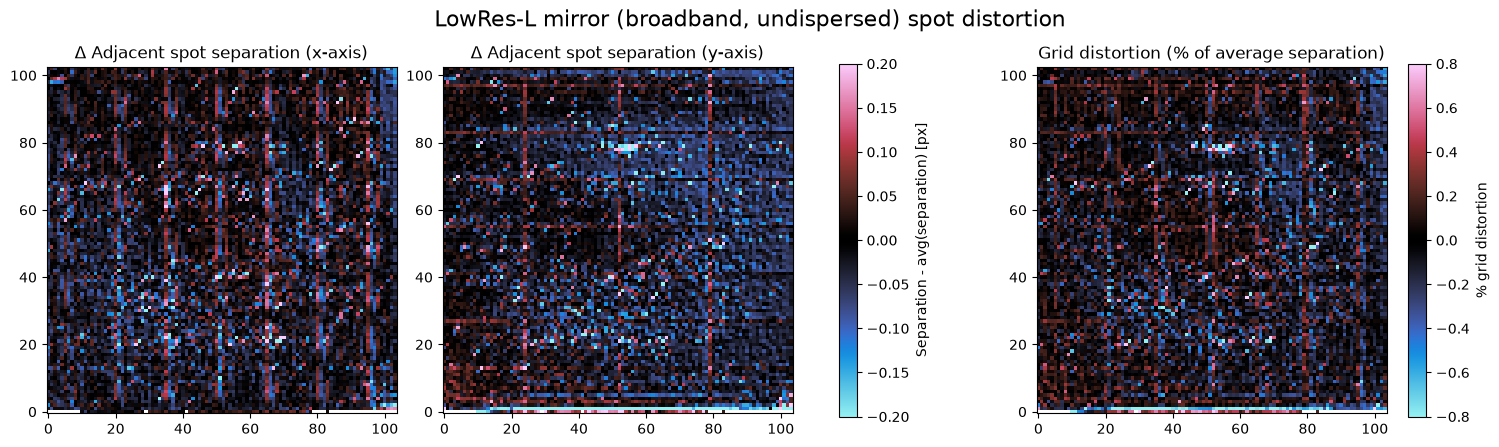}
        \caption{Optical distortion measured for low-res mirror spots (a mirror instead of a disperser is inserted into the beam, creating broadband, undispersed lenslet spots on the detector). Left two panels: change in separation between adjacent lenslet spots across the FOV, measured along either the x- or y-axis. Right panel: grid distortion, measured as a percent of the nominal spot-to-spot separation. A grid pattern is visible, which is imparted by the lenslet array itself. }
        \label{fig:distortion-mirror}
    \end{subfigure}
    \begin{subfigure}{1.0\textwidth}  
        \includegraphics{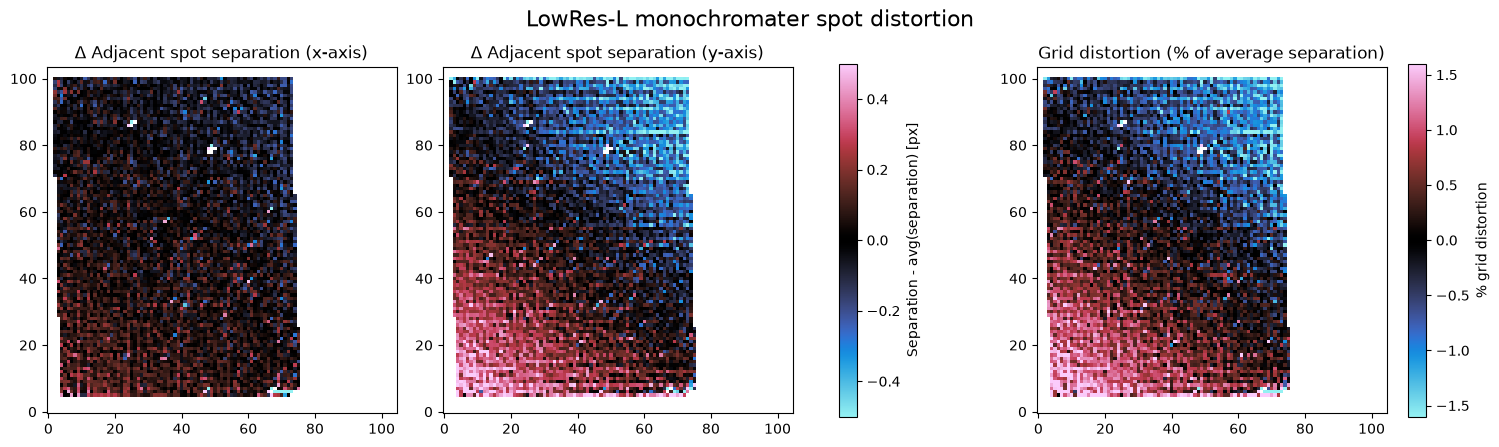}
        \caption{Optical distortion measured for low-res single-wavelength spaxels; while the cross-hatching pattern from the lenslet array is visible, monochromater spot distortion is dominated by anamorphic magnification.}
        \label{fig:distortion-mono}
    \end{subfigure}
    \caption{Optical distortion of LowRes-L lenslet spots. The SCALES DRP accounts for grid distortion of lenslet spots during datacube reconstruction, meaning static distortion is of no concern for SCALES users.}
    \label{fig:distortion}
\end{figure*}

We present measurements of lenslet spot distortion for the low-res IFU in Figure \ref{fig:distortion}. Each lenslet in the array is 341 x 341$\mu$m, which implies a separation of 18.94px between adjacent lenslet spots on the detector. However, a small amount of differential magnification across the FOV is inherent in the spectrograph design, leading to 18.9px lateral and 19.1px vertical spot separations, with small variations across the field of view. Here we take grid distortion to be deviations from the nominal ($\Delta$x=18.9px, $\Delta$y=19.1) separation between adjacent pixels. Any distortion of the lenslet spots in addition to this differential magnification is static, imparted by the spectrograph optics downstream of the lenslet array as well as any manufacturing defects in the lenslet array itself. 

Figure \ref{fig:distortion-mirror} shows measured distortion of the undispersed LowRes-L mirror spots. The left two panels show distortion in pixels in the x and y directions respectively; the right panel shows grid distortion as a percentage of the nominal xy separation between pixels. We find distortion to be extremely low, with measured shifts of $\leq$0.2px along either axis and maximum grid distortion of $\leq$0.8\% across the low-res FOV. Distortion of undispersed spots is dominated by a grid pattern imparted by the structure of the lenslet array itself – visual inspection of the array shows that lenslets are grouped into 15x15 clusters. 

Figure \ref{fig:distortion-mono} shows measured distortion of the dispersed single-wavelength LowRes-L spots. Misalignment of the TelSim resulted in uneven illumination during datataking; the data were truncated to a subset of the low-res FOV so that only regions of sufficient SNR were included in this analysis. The truncated FOV for the dispersed data is approximately registered with the undispersed mirror spots for easier visual comparison. While the grid pattern from the lenslet array itself is still visible, distortion of dispersed monochromater spots is dominated (primarily in the y-direction) by anamorphic magnification, roughly aligned with the measured direction of dispersion for the L-prism (71.58$^{\circ}$ from the +x axis).


\subsubsection{Enslitted energy}
\label{subsec:ee}

\begin{figure}[ht]
     \centering
     \begin{subfigure}[b]{1.0\textwidth}
         \centering
         \includegraphics[width=\textwidth]{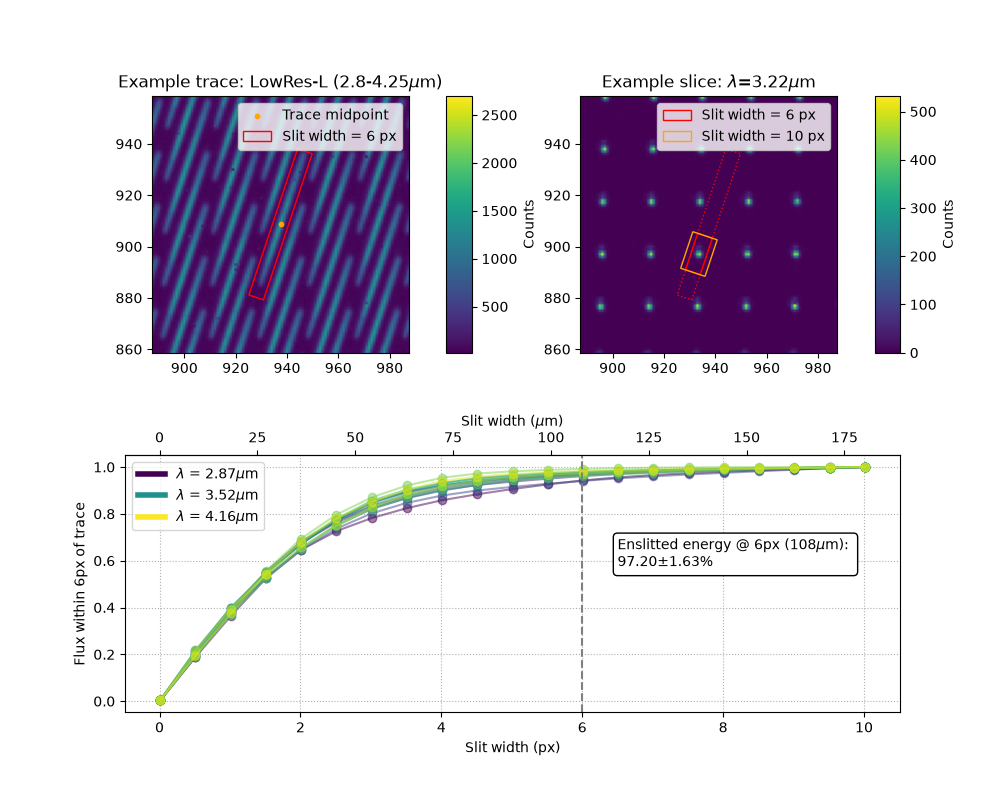}
    \end{subfigure}
        \caption{Enslitted energy measurement for a single spaxel in L-band (2.8-4.25$\mu$m); results are similar for spaxels across the full FOV.}
        \label{fig:ee}
\end{figure}

We measure the enslitted energy of the low-res IFU, which is a measure of the concentration of flux from each lenslet spot. We measure the amount of flux from each lenslet spot that falls within a rectangular slit of width 6px, i.e. the separation between adjacent traces; flux that falls outside of this slit width amounts to cross-talk, and bleeds into adjacent lenslet spots. Figure \ref{fig:ee} shows enslitted energy measurement for a LowRes-L monochromater datacube taken during CD3. The top left panel shows the full trace (created by summing the datacube together along the wavelength axis), with the 6px-wide rectangular aperture plotted for visual reference. The top right panel shows a single slice of the datacube, where the single-wavelength monochromater light forms spots on the detector after being dispersed. A smaller rectangular aperture is placed directly over the lenslet spot (annotated in orange), and flux is measured over this aperture as its width is varied from 0-10px. We assume that all spot flux is captured by a 10px-wide aperture. We find that enslitted energy is typically $\gtrsim$97\% averaged across wavelengths and across the low-res FOV, which suggests $\sim$3\% crosstalk. This is in line with optical simulations conducted for FDR which predicted $\sim$2\% crosstalk. More precise crosstalk characterization will be conducted during commissioning. We note that crosstalk is slightly higher at longer wavelengths, which is expected due to the wavelength dependence of PSF size.

\subsection{IFU image quality}
\label{subsec:image-quality}

\begin{figure}[h!]
     \centering
     \begin{subfigure}[b]{0.31\textwidth}
         \centering
         \captionsetup{width=0.9\textwidth}
         \includegraphics[width=1.0\textwidth]{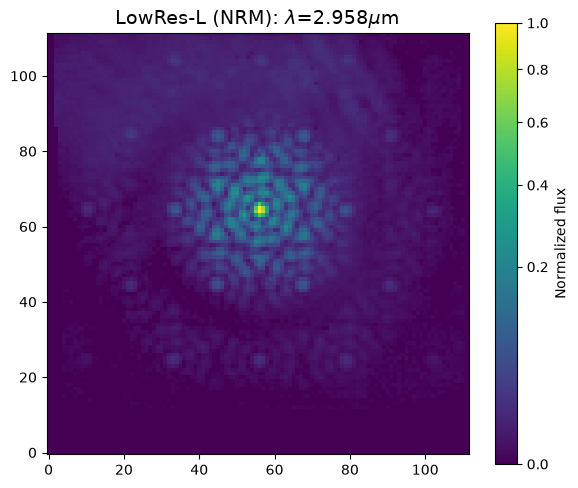}
         \caption{LowRes-L (2.8-4.25$\mu$m)  PSF from a non-redundant mask (NRM).}
     \end{subfigure}
     \begin{subfigure}[b]{0.31\textwidth}
         \centering
         \captionsetup{width=0.9\textwidth}
         \includegraphics[width=1.0\textwidth]{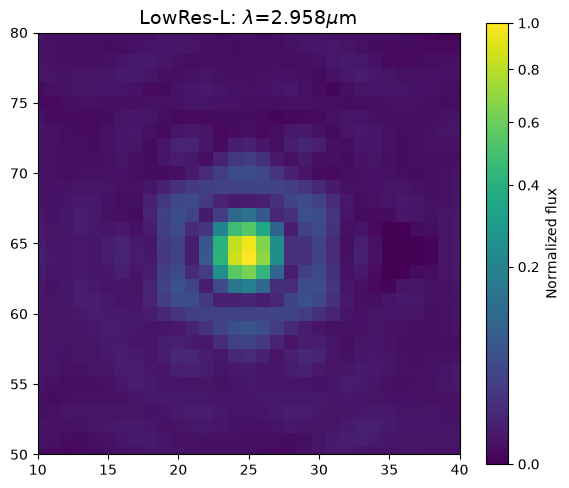}
         \caption{LowRes-L (2.8-4.25$\mu$m) PSF cutout from a 25$\mu$m pinhole at the coronagraph focal plane.}
     \end{subfigure}
     \begin{subfigure}[b]{0.31\textwidth}
         \centering
         \captionsetup{width=0.9\textwidth}
         \includegraphics[width=1.0\textwidth]{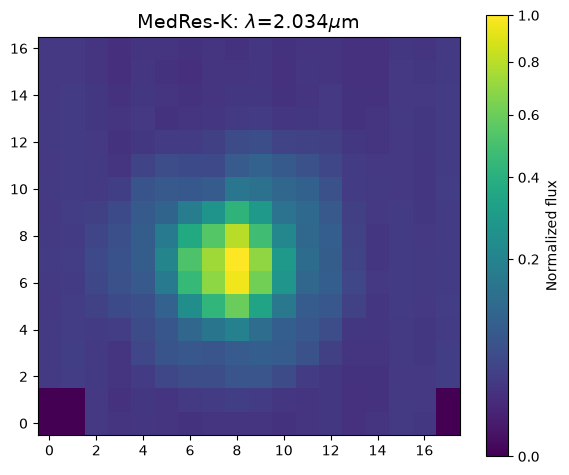}
         \caption{MedRes-K (1.9-2.5$\mu$m) PSF, also from the pinhole.}
         \vspace{3.8mm}
     \end{subfigure}
     \hfill
        \caption{SCALES IFU PSFs. These should not be taken as representations of end-to-end performance, given the constraints of lab testing, but as initial assessments of image quality.}
        \label{fig:psf}
\end{figure}

Finally, we assess the image quality of the low- and med-res IFU. These are not end-to-end measurements – with the UCSC cleanroom test configuration, it is not possible to inject a perfect point source into SCALES, so approximations are made here for the purposes of preliminary analysis. Figure \ref{fig:psf} shows three quasi-point-source images taken during CD5: the left two panels show low-res IFU PSFs created by one of SCALES' non-redundant masks (see Lach et al. 2026 \cite{mackenzie_r_lach_scales_2026}) and a 25$\mu$m pinhole mask inserted into the coronagraph focal plane; the right panel shows a med-res PSF also of the 25$\mu$m pinhole. The low- and med-res IFU PSFs are visually very clean, and the low-res PSF is diffraction-limited. 

The pinhole mask used to create an artificial point-source is reflective, which introduces structured background noise in the vicinity of the PSF and makes Strehl difficult to measure, especially at longer wavelengths where the pinhole background looks worse. Taking wavelengths with visually less background contamination, we measure a Strehl of 93.2\% for the low-res pinhole images at $\lambda$=3.260$\mu$m, but the true Strehl is likely higher. Further work will be carried out during commissioning to more precisely characterize SCALES' end-to-end image quality and wavefront error. 


\appendix    

\acknowledgments 
 
This work was enabled by funding from the NSF, including the NSF Graduate Research Fellowship Program (NSF GRFP). Major support for the SCALES project has also been provided from the Heising-Simons Foundation, the Mt. Cuba Astronomical Foundation, and the Alfred P. Sloan Foundation. We are also grateful to the Robinson family and other private supporters, whose generosity has been instrumental in making this work possible. This work benefited from the 2025 Exoplanet Summer Program in the Other Worlds Laboratory (OWL) at the University of California, Santa Cruz, a program funded by the Heising-Simons Foundation and NASA. We would like to thank everyone at Keck Observatory who have generously lent their time and expertise.

\printbibliography

\end{document}